\documentclass[manuscript]{acmart}

\usepackage{graphicx}

\AtBeginDocument{%
  \providecommand\BibTeX{{%
    \normalfont B\kern-0.5em{\scshape i\kern-0.25em b}\kern-0.8em\TeX}}}

\setcopyright{acmcopyright}
\copyrightyear{2022}
\acmYear{2022}
\acmDOI{XX.XXXX/XXXXXXX.XXXXXXX}

\acmConference[CHI 2022]{New Orleans '22: ACM Conference on Human Factors in Computing Systems}{April 30 -- May 6, 2022}{New Orleans, LA}
\acmBooktitle{CHI 2022: Workshop on Trust and Reliance in AI-Human Teams,
  April 30 -- May 6, 2022, New Orleans, LA}
\acmPrice{15.00}
\acmISBN{XXX-X-XXXX-XXXX-X/XX/XX}

\begin{document}

\title[Distinguishing Trust and Reliance in XAI]{Trust and Reliance in XAI - Distinguishing Between Attitudinal and Behavioral Measures}

\author{Nicolas Scharowski}
\authornote{Corresponding Author.}
\email{nicolas.scharowski@unibas.ch}
\orcid{0000-0001-5983-346X}
\affiliation{
  \institution{Center for General Psychology and Methodology, University of Basel}
  \streetaddress{Missionsstrasse 62a}
  \city{Basel}
  \postcode{CH-4055}
  \country{Switzerland}
}

\author{Sebastian A. C. Perrig}
\email{sebastian.perrig@unibas.ch}
\orcid{0000-0002-4301-8206}
\affiliation{
  \institution{Center for General Psychology and Methodology, University of Basel}
  \city{Basel}
  \country{Switzerland}
}

\author{Nick von Felten}
\email{nick.vonfelten@unibas.ch}
\orcid{0000-0003-0278-9896}
\affiliation{
  \institution{Center for General Psychology and Methodology, University of Basel}
  \city{Basel}
  \country{Switzerland}
}

\author{Florian Brühlmann}
\email{florian.bruehlmann@unibas.ch}
\orcid{0000-0001-8945-3273}
\affiliation{
  \institution{Center for General Psychology and Methodology, University of Basel}
  \city{Basel}
  \country{Switzerland}
}

\renewcommand{\shortauthors}{Scharowski et al.}

\begin{abstract}

Trust is often cited as an essential criterion for the effective use and real-world deployment of AI. Researchers argue that AI should be more transparent to increase trust, making transparency one of the main goals of XAI. Nevertheless, empirical research on this topic is inconclusive regarding the effect of transparency on trust. An explanation for this ambiguity could be that trust is operationalized differently within XAI. In this position paper, we advocate for a clear distinction between behavioral (objective) measures of reliance and attitudinal (subjective) measures of trust. However, researchers sometimes appear to use behavioral measures when intending to capture \emph{trust}, although attitudinal measures would be more appropriate. Based on past research, we emphasize that there are sound theoretical reasons to keep trust and reliance separate. Properly distinguishing these two concepts provides a more comprehensive understanding of how transparency affects trust and reliance, benefiting future XAI research.

\end{abstract}



\maketitle

\section{Introduction}

Recent advances in artificial intelligence (AI) and its applicability in various everyday applications (e.g., video surveillance, email spam filtering, and product recommendations) have prompted a demand for more transparent AI to mitigate the potentially negative consequences of opaque AI \citep{FAccT, CHI21}. This call for transparent AI has led to the multidisciplinary research field of explainable artificial intelligence (XAI), which explores methods and models that make the behaviors, predictions, and decisions of AI transparent and understandable to humans \citep{liao2020questioning, Lipton.10.06.2016}. There are different rationales for transparent AI, one of which is the assumption that transparency contributes toward building a relationship of trust between humans and AI \citep{Stephanidis.2019} and more reliance on AI \citep{hoff2015trust}. Past research has shown that people tend to rely on automation they trust and reject automation they distrust \citep{lee2004trust}. This makes trust particularly relevant in the misuse (overreliance \citep{Parasuraman1997}) and disuse (neglect or underutilization \citep{Parasuraman1997}) of automation and AI \citep{Stephanidis.2019, Yu.2017, hoff2015trust}, rendering trust a key property of the interaction between users and AI \citep{jacovi2021formalizing}.

However, there is mixed empirical evidence about whether transparency is indeed increasing trust \citep{Ehsan.11.01.2019, poursabzi2018manipulating, cheng2019explaining, Cramer, nothdurft2013impact, Zhang2020}. We argue that one possible reason for this ambiguity could be that within empirical XAI research, trust is measured in different ways, and often there does not appear to be a clear distinction between trust and related concepts such as reliance in the XAI community. To provide two examples from empirical work: \citeauthor{Lai2019} \citep{Lai2019} conducted an experiment, asking end-users to decide whether a hotel review was genuine or deceptive. They illustrated that heatmaps of relevant instances and example-based explanations improved human performance and increased the trust humans place in the predictions of AI. \citeauthor{Lai2019} defined trust as the percentage of instances for which humans relied on the machine prediction. In contrast, \citeauthor{cheng2019explaining} \citep{cheng2019explaining} conducted an experiment where participants used different UI interfaces to comprehend an algorithm's decision for university admissions. They showed that revealing the inner workings of an algorithm can improve users' comprehension and found that users' trust, assessed by a survey scale, was not affected by the explanation interface. These two examples illustrate how trust can be thought of and measured differently in XAI research. Both studies intended to measure \emph{trust}, but their respective operationalization of it varied substantially. \citeauthor{Lai2019} \citep{Lai2019} opted for a \emph{behavioral measurement} of trust (percentage of reliance instances) while \citeauthor{cheng2019explaining} \citep{cheng2019explaining} chose an \emph{attitudinal measurement} of trust (scale-based measurement).

Researchers' choice of measures has implications for the variable of interest, and we will demonstrate that there are compelling theoretical reasons to distinguish between trust and reliance. We argue that a lack of distinction between behavioral and attitudinal measures leads to misunderstandings regarding what is actually being investigated in a study - trust or reliance. This misunderstanding could be a reason for the inconclusive findings in current XAI literature regarding the effect of transparency. For this reason, a rigorous distinction between trust and related concepts like reliance seem warranted. Although the two mentioned studies \citep{Lai2019, cheng2019explaining} are merely illustrative, they indicate conceptual inconsistencies in current XAI research that needs to be addressed and provide a research opportunity to better define and formalize key concepts of XAI such as trust and reliance. We encourage a comprehensive literature review of past empirical work to better understand the extent to which such conceptual inconsistencies exist in current XAI literature. The purpose of this position paper is to emphasize the importance of adequately distinguishing between trust and reliance in measurement and provide theoretical reasons to make this distinction. We will further demonstrate that this measurement indifference has implications for the conclusions derived from XAI research. If research properly considers and distinguishes different ways of measuring trust and related concepts, this can bring great potential for a more comprehensive understanding of XAI.

\section{The distinction between trust as an attitude and reliance as a behavior}

\citeauthor{lee2004trust} \citep{lee2004trust} defined trust in automation as "the attitude that an agent will help achieve an individual's goals in a situation characterized by uncertainty and vulnerability." \citep[p. 6]{lee2004trust} This definition encapsulates the notion of uncertainty, vulnerability, and risk, a necessary prerequisite for human trust in AI \citep{jacovi2021formalizing} and is consistent with the notion of vulnerability emphasized in the most widely used and accepted definition of trust \citep{rousseau1998not} by \citeauthor{mayer1995integrative} \citep{mayer1995integrative}.

Defining trust as an \emph{attitude} suggests that it should be regarded as a psychological construct. In this context, the term construct refers to unobservable features, such as psychological traits or abilities of people \cite{hopkins1998educational}. Psychological constructs are typically measured by survey scales, sometimes referred to as questionnaires or rating scales. The data collected using survey scales is inherently subjective, given that it reflects participants' own perspectives. At their core, survey scales are about using a series of items to measure a construct of interest to a researcher \cite{devellis2017scale}. Related to XAI research, a researcher could be interested in a user's \emph{subjective} trust in an automated system (i.e., the construct "trust in automation") and have users respond to the 12 items of the survey scale by \citet{jian2000foundations}. Combining the responses to the items of the scale (e.g., by forming a total score or calculating the mean value) provides the researcher with a number reflecting users' \emph{subjective level of trust} toward the system. This number can then be used for statistical analysis during the research process and compared across research if the specific scale is used more than once.

\citeauthor{lee2004trust} \citep{lee2004trust} thus defined trust as an \textit{attitude} and distinguished it from reliance as a \textit{behavior}. In contrast to trust as a psychological construct, \emph{behavior} (e.g., reliance) is directly observable and therefore open to more objective approaches of measurement. Based on \citeauthor{lee2004trust} we define \emph{reliance} on a system as \emph{a user's behavior that follows from the advice of the system}. In their work on metrics for XAI, \citeauthor{hoffman2018metrics} \citep{hoffman2018metrics} make a similar distinction when they differentiate between trusting a machine's output and following its advice. Given this definition, one possibility to measure reliance is "weight of advice" (WOA), stemming from the literature on advice taking \citep{harvey1997taking}. WOA measures the degree to which people change their behavior and move their initial estimate (e.g., a price estimate) towards an (AI) advice to form a final estimate. While WOA has been used in the past by researchers in XAI as a more objective trust measurement \citep{poursabzi2018manipulating, Logg.2019, Mucha}, it has never been explicitly referred to as reliance and thus clearly differentiated from trust to the best of our knowledge. 

Given these arguments, a proposed distinction between trust and reliance relates to subjective and objective measurement approaches. This differentiation between subjective and objective measures of trust in XAI was addressed by \citeauthor{mohseni2018multidisciplinary} \citep{mohseni2018multidisciplinary}. They pointed out that subjective trust measures in XAI include interviews, survey scales, and self-reports via questionnaires. More recently, \citeauthor{buccinca2020proxy} \citep{buccinca2020proxy} pointed out that subjective measures of trust have been a focal point in evaluating AI. For objective measures of trust, \citeauthor{mohseni2018multidisciplinary} \citep{mohseni2018multidisciplinary} proposed user's system competence, understanding, and users' compliance (i.e., reliance) with a system. Given these clarifications, it is evident that \citeauthor{Lai2019} \citep{Lai2019} and \citeauthor{cheng2019explaining} \citep{cheng2019explaining} made different choices of method (objective behavioral measurement vs. subjective scale-based measurement) while intending to capture the same variable of interest (i.e., trust). 

However, the choice of measurement has implications on the variable of interest. Assuming that trust as an attitude is a psychological construct, behavioral measures are less adequate to capture its subjective nature, and scale-based approaches (e.g., \citeauthor{jian2000foundations} \citep{jian2000foundations}) are more appropriate. Conversely, attitudinal measures may be inadequate to reflect the objective constituents of reliance as a behavior. This indicates that XAI research could be at risk of confusion between the variable of interest and their respective attitudinal and behavioral measures. This lack of clarity regarding what is measured has implications not only for the data gathered with a particular method but also for the conclusions derived from said data, and thus the validity of a study as a whole \citep{mackenzie2003dangers}. We argue that this ambiguity could hinder the comparison of empirical work in XAI since researchers may have the same variable of interest in mind (e.g., trust) but miss each other’s work or implications because of different measurement choices. Rigorously distinguishing between trust and reliance also offers a more comprehensive understanding of the effects of transparency and is, therefore, an opportunity for the XAI community.

\section{Discussion}

We argue that the above-mentioned conceptual considerations have been insufficiently considered in past XAI research and emphasize that researchers need to differentiate between trust and reliance, further accentuating the importance of this distinction proposed by \citeauthor{lee2004trust} \citep{lee2004trust}. The conceptual distinction between trust and reliance carries significant implications for XAI evaluation and uncovers two potential challenges. First, if researchers only assess trust as an attitude via questionnaires, they could falsely assume that people will (or will not) rely on an AI system. Second, if only behavior (i.e., reliance) is measured, researchers might incorrectly deduce that people necessarily trust (or not trust) the system in question. Consequently, researchers and practitioners have to decide which variables of interest to consider and investigate when evaluating AI. We think it is useful to conceptualize trust as a psychological construct that is associated with reliance but does not fully determine it \citep{lee2004trust}. While trust is considered the main factor for human reliance on automation, other factors (e.g., workload, self-confidence, and task-associated risks) are likewise important \citep{lee2004trust, hussein2020trust} and whether trust translates into reliance is more nuanced than often assumed and depends on an interaction between the operator, the automation, and the situation \citep{lee2004trust, korber2018theoretical}. 

We propose that it is sensible to consider trust and reliance to be conceptually distinct, because attitudes and behavior do not share a deterministic but a probabilistic relationship \citep{fishbein1980predicting}. While researchers have shown that attitudinal judgments have an impact on people's reliance on automated systems \citep{Cramer, merritt2011affective}, trust does not necessarily lead to more reliance. Even if an AI system is trusted, reliance must not follow \citep{kirlik1993modeling, korber2018theoretical}, and people may claim to trust an AI system, yet behave in a way that suggests they do not \citep{miller2016behavioral}. The empirical findings of \citeauthor{dzindolet2003role} \citep{dzindolet2003role} support this argument: although some automated decision aids were rated as more trustworthy than others, all were equally likely to be relied upon. Multiple models of the relationship between trust and reliance have been proposed \citep{hoff2015trust, lee2004trust, vandongen2013framework, chancey2017trust, schaefer2016meta} and it has been hypothesized that trust mediates between human reliance and reliability of automation \citep{lee2004trust, Parasuraman1997}. This relationship has recently been empirically supported \citep{hussein2020trust}. Research on XAI should take such models as a theoretical basis for their empirical work and refine them whenever necessary in the context of AI to advance the field with a rigorous science underlying it.

While we have provided a theoretical justification for a distinction between trust and reliance, others have provided more philosophical arguments, suggesting that the use of the term trust is not appropriate and sensible in the context of technology \citep{jacovi2021formalizing, hartmann2020vertrauen, baier1986trust}. Because trust requires the acceptance of vulnerability, but objects have no understanding about this vulnerability, they cannot intentionally betray our trust, we can only rely on them. Therefore, so the argument, it might be unreasonable to talk about trust when referring to interactions between humans and AI, and instead, it is a matter of reliance. \citeauthor{jacovi2021formalizing} \citep{jacovi2021formalizing} argued that using the term trust is appropriate when attributing AI with intent and that in the case of non-anthropomorphized AI, the term reliance should be considered. Given this philosophical perspective, there appears to be potential for future research distinguishing between trust and reliance, both with anthropomorphized and non-anthropomorphized AI. Such research could then provide empirical evidence, adding to the philosophical discussion raised by \citeauthor{baier1986trust} \citep{baier1986trust} and others \citep{hartmann2020vertrauen, jacovi2021formalizing}.

\citeauthor{Arrieta.2019} \citep{Arrieta.2019} emphasized that only agreed-upon metrics and their respective measurements allow for meaningful comparisons in XAI research and that without such consistency, empirical claims are ambiguous and do not provide a solid foundation for future XAI research. Therefore, it seems particularly important that researchers carefully differentiate \emph{attitude} from \emph{behavior} and choose appropriate measures for human-centred AI accordingly. Future research could conduct a literature review of how measures are used in the current XAI literature to understand better the extent to which conceptual inconsistencies exist between subjective attitudinal and objective behavioral measures. Such work could further develop a framework to guide the selection of measures of trust and reliance, depending on different research purposes. A way forward could be measuring both, trust \emph{and} reliance and recognizing theoretical models on their relationship that have been proposed by past research \citep{hoff2015trust, lee2004trust, vandongen2013framework, chancey2017trust, schaefer2016meta}. Given that AI is increasingly applied to make critical decisions with far-reaching consequences, selecting appropriate measurements for clearly defined concepts is of paramount importance in XAI. 

\bibliographystyle{ACM-Reference-Format}
\bibliography{sample}

\end{document}